\documentclass[12pt,a4paper]{article}
\usepackage{amsmath,amssymb,amsthm}
\usepackage[margin=1.0in]{geometry}
\usepackage{cite}
\usepackage[colorlinks=true
,urlcolor=blue
,anchorcolor=blue
,citecolor=blue
,filecolor=blue
,linkcolor=blue
,menucolor=blue
,pagecolor=blue
,linktocpage=true
,pdfproducer=medialab
,pdfa=true
]{hyperref}
\numberwithin{equation}{section}

\def\a{\alpha}

\def\g{\gamma}

\def\e{\epsilon}

\def\z{\zeta}

\def\k{\kappa}
\def\l{\lambda}
\def\m{\mu}
\def\n{\nu}

\def\r{\rho}

\def\s{\sigma}

\def\ph{\phi}

\def\be{\begin{equation}}
\def\ee{\end{equation}}
\def\bea{\begin{eqnarray}}
\def\eea{\end{eqnarray}}

\def\pa{\partial}

\def\td{\tilde}

\def\lp{\left(}
\def\rp{\right)}
\def\ls{\left[}
\def\rs{\right]}
\def\nn{\nonumber}
\def\ie{{\it i.e., }}

\makeatletter
\renewcommand\section{\@startsection {section}{1}{\z@}%
                                   {-3.5ex \@plus -1ex \@minus -.2ex}
                                   {2.3ex \@plus.2ex}%
                                   {\normalfont\large\bfseries}}
\renewcommand\subsection{\@startsection{subsection}{2}{\z@}%
                                     {-3.25ex\@plus -1ex \@minus -.2ex}%
                                     {1.5ex \@plus .2ex}%
                                     {\normalfont\bfseries}}
\makeatother

\begin{document}

\begin{center}
\addtolength{\baselineskip}{.5mm}
\thispagestyle{empty}
\begin{flushright}
\today \\
\end{flushright}

\vspace{20mm}

{\Large  \bf Two Ramond-Ramond corrections to type II supergravity via field-theory amplitude}
\\[15mm]
{Hamid R. Bakhtiarizadeh\footnote{bakhtiarizadeh@sirjantech.ac.ir}}
\\[5mm]
{\it Department of Physics, Sirjan University of Technology, Sirjan, Iran}

\vspace{20mm}

{\bf  Abstract}
\end{center}

Motivated by the standard form of string-theory amplitude, we calculate the field-theory amplitude to complete the higher-derivative terms in type II supergravity theories in their conventional form. We derive explicitly the $ O(\a'^3) $ interactions for the RR (Ramond-Ramond) fields with graviton, B-field and dilaton in the low-energy effective action of type II superstrings. We check our results by comparison with previous works that have been done by the other methods, and find an exact agreement.

\vfill
\newpage


\section{Introduction}

Higher-derivative corrections to string theories and M-theory are significantly studied in various ways: string amplitude \cite{Gross:1986iv,Kiritsis:1997em,Russo:1997mk,Green:1997di,Green:1997as,Anguelova:2004pg,Stieberger:2009hq}, non-linear sigma model \cite{Gross:1986mw,Grisaru:1986px}, superfield and noether's method \cite{deRoo:1992zp,deRoo:1992sm,Rajaraman:2005up,Paulos:2008tn,Cremmer:1978km,Hyakutake:2006aq,Rajaraman:2005ag,Peeters:2000qj}, duality completion \cite{Garousi:2013nfw,Godazgar:2013bja,Liu:2013dna,Garousi:2017fbe} and so on. Each of these approaches have been employed in different formalisms such as: RNS (Ramond-Neveu-Schwarz) \cite{Peeters:2003pv,Bakhtiarizadeh:2013zia,Bakhtiarizadeh:2015exa}, GS (Green-Schwarz) \cite{Liu:2013dna,Nilsson:1981bn,Howe:1983sra,Peeters:2000qj} and pure-spinor \cite{Policastro:2006vt} formalisms to determine the higher-order terms.

For many purposes, it is enough to use only the lowest-order terms in the theory, but there are some situations for which one must go beyond the lowest-order supergravity actions and higher-order corrections play an important role. For example, the origin of induced Einstein-Hilbert terms are traced to $ R^4 $ couplings in ten dimensions \cite{Antoniadis:1997eg,Antoniadis:2002tr}. Furthermore, in black-hole and black-brane physics, considering higher-derivative terms leads to modifications of the thermodynamics \cite{Iyer:1994ys,deHaro:2003zd,Kraus:2005vz}. Another important set of such applications can be found in the context of the gauge/gravity duality \cite{Gubser:1998nz,Pawelczyk:1998pb,Frolov:2001xr}. There are a vast number of applications of higher-derivative terms, but the mentioned examples are sufficient to illustrate the importance of having a good understanding of higher-derivative terms \cite{Peeters:2003pv}.

It is known that the low-energy effective action of superstring theory is given by supergravity describing only the interactions of massless modes in the string-theory spectrum. This can be shown explicitly by calculating the field-theory amplitudes of massless states \cite{Sannan:1986tz}. One must go beyond this low-energy limit to capture truly stringy behavior. A significant information about string and M-theory can be extracted from the corresponding low-energy effective actions, in particular once one considers corrections that go beyond the leading order. Subleading terms in type II effective actions start at order $ \a'^3 $ or eight-derivative level. In order to determine the structure of higher-derivative terms, we will calculate the scattering amplitude of massless states.

As mentioned above, scattering amplitudes of massless states in superstring theory include corrections to their corresponding low-energy effective actions. These terms contain $ \a' $ corrections to the supergravity which arise due to the length of the fundamental string $ \ell_s $ and string coupling constant $ g_s $ which correspond to string quantum corrections in spacetime. In type IIA superstring theory, one of those corrections first obtained at the tree level from four-graviton scattering amplitude as well
as from the $ \s $-model beta function approach which is written by
\bea
{\cal L}_{R^4}=-\frac{e^{-2\ph}}{2\k_{10}^{2}}\frac{\ell_{s}^6}{2^8.4!}\z(3)\lp t_8t_8R^4+\frac{1}{4.2!}\e_{10} \e_{10} R^4\rp.
\eea
Here $ 2\k_{10}^{2}= (2\pi)^7 \ell_s^8 g_s^2 $, $ t_8 $ is a product of four Kronecker deltas, and $ t_8t_8R^4 $ denotes an abbreviation of a product of two $ t_8 $ tensors and four Riemann curvature tensors.

The contents of this paper is organized as follows: In section \ref{brII}, we briefly review type II supergravity theories and fix our conventions and notations. In section \ref{ftamp}, we establish the formalism and explain the procedure needed to find the necessary Feynman rules for the processes we want to compute. By employing these rules we then calculate the tree-level four-point amplitude for two RR-two NSNS scattering to find higher-derivative corrections to type II supergravities in their conventional form up to the $ R^4 $ terms. We restrict our attention to different external NSNS states, namely the metric, antisymmetric tensor and dilaton. Finally, we compare our results with previous works and find an exact agreement.

\section{Brief review of type II supergravity}\label{brII}

We begin by reviewing the low-energy effective actions of both type II superstring theories \cite{Becker:2007}. These are supergravity theories that describe interactions of the massless fields in the string-theory spectrum. The action in the Einstein frame for the type IIA supergravity is given by
\bea
S_{\rm IIA}&=&\frac{1}{2\kappa^2} \int d^{10}x \sqrt{-G} \left(R-\frac{1}{2} \pa_{\mu} \Phi \pa^{\mu} \Phi-\frac{1}{2} e^{-\Phi} |H|^2-\frac{1}{2} \sum_{n=2,4}e^{\frac{5-n}{2}\Phi} |\tilde{F}^{(n)}|^2\right)\nonumber\\
&&-\frac{1}{4\kappa^2} \int\, B\wedge dC^{(3)}\wedge dC^{(3)},\label{IIA}
\eea
where $ R $ is the scalar curvature, $ \Phi $ is the dilaton field and $H$ is the B-field strength  $H=dB$. The RR field strengths are defined in terms of RR potentials as $\tilde{F}^{(2)}=dC^{(1)}$ and $\tilde{F}^{(4)}=dC^{(3)}-   H\wedge C^{(1)}$. The above action is the   reduction of  11-dimensional supergravity on manifold $R^{1,9}\times S^1$. By a Weyl rescaling of the metric, this action can be transformed to the Einstein frame in which the Einstein term has the conventional form\footnote{The Einstein frame is related to string frame by $ g_{\m\n}^{\rm E} = e^{-\Phi/2} g_{\m\n} $.}. It is also contains three distinct types of terms. The first three terms in the first line involve NSNS fields, which are common to both type II superstring theories. The last term contains RR fields and altogether are named kinetic terms. The second line is called Chern-Simons term.

In the type IIB supergravity, the presence of the self-dual five-form introduces a significant complication for writing down a classical action for type IIB supergravity. On the other words, it is hard to formulate the action in a manifestly covariant form. There are several different ways of dealing with this problem. One of them is to find an action which reproduce the super-symmetric equations of motion when the self-duality condition is imposed by hand. The type IIB supergravity action in the Einstein frame is given as
\bea
S_{\rm IIB}&=&\frac{1}{2\kappa^2} \int d^{10}x \sqrt{-G} \left(R-\frac{1}{2} \pa_{\mu} \Phi \pa^{\mu} \Phi-\frac{1}{2} e^{-\Phi} |H|^2-\frac{1}{2\alpha} \sum_{n=1,3,5} e^{\frac{5-n}{2}\Phi} |\tilde{F}^{(n)}|^2\right)\nonumber\\
&&-\frac{1}{4\kappa^2} \int\,  H\wedge dC^{(2)}\wedge C^{(4)},\label{IIB}
\eea
where $\alpha=1$ for $n=1,3$ and  $\alpha=2$ for $n=5$. The RR field strengths here are defined as $\tilde{F}^{(1)}=dC^{(0)}$,   $\tilde{F}^{(3)}=dC^{(2)}- H C^{(0)}$ and 
\bea
\tilde{F}^{(5)}=dC^{(4)}-\frac{1}{2} C^{(2)} \wedge H+ \frac{1}{2} B \wedge dC^{(2)}, \label{tildeF5}
\eea
respectively. The self-duality condition that must be imposed  in the equations of motion by hand, is given by
\bea
\tilde{F}^{(5)}=\star \tilde{F}^{(5)} . \label{selfdual}
\eea
This condition has to be imposed as an extra constraint, manually. Without that, one can not find any consistency between the field-theory and string-theory S-matrix elements.
\section{Field-theory amplitude}\label{ftamp}

Having had the supergravity actions, one can easily read different vertices and propagators and accordingly calculate the Feynman amplitude of two RR-two NSNS massless states. To that end, suppose that the massless fields are small perturbations around the flat background, \ie 
\bea
g_{\mu\nu}=\eta_{\mu\nu}+2\kappa h_{\mu\nu};\,B^{(2)}=2\kappa b^{(2)}\,;\,\,
\Phi\,=\,\phi_0+\sqrt{2}\kappa \phi\,;\,\,{C}^{(n)}\,=\,2\kappa c^{(n)}, \label{perturb}
\eea
where $ \k $ is the gravitational coupling constant. Substituting these perturbation expansions into the supergravity actions (\ref{IIA}) and (\ref{IIB}), and expanding them up to cubic and quartic powers in $ \k $, one obtains three-point and four-point interactions, respectively. The amplitude of two RR-two NSNS fields typically has the following form 
\bea
{\cal A}&=& A_s+A_u+A_t+A_c\nn\\&=&\frac{{\cal A}_s}{s}+\frac{{\cal A}_u}{u}+\frac{{\cal A}_t}{t}+{\cal A}_c\nn\\&=&\frac{1}{sut}\lp tu{\cal A}_s+st {\cal A}_u+su{\cal A}_t+sut{\cal A}_c\rp ,\label{ampf}
\eea
where $ {\cal A}_s, {\cal A}_u $ and $ {\cal A}_t $ are the amplitudes in $ s $, $ u $ and $ t $ channels, respectively, and $ {\cal A}_c $ is the contribution of contact terms which must be added in order to the total scattering amplitude to be gauge invariant. The Mandelstam variables are defined as: $ s = -4 \a' k_1\cdot k_2 $, $ u = -4 \a' k_1\cdot k_3 $, $ t = -4 \a' k_2\cdot k_3 $, and satisfy the identity $ s + t + u = 0 $.

What motivates us to write the field-theory amplitude in the above form originates from the general form of string-theory amplitude. The structure of string-theory amplitude consists of the well-known Gamma functions in terms of Mandelstam variables multiplied by a kinematic factor $ \cal K $ \cite{Gross:1986iv,Schwarz:1982jn}. To find the couplings which are reproduced by the amplitude in string-theory side, it is convenient to expand the Gamma functions at low energy, \ie  
\bea
\frac{\Gamma(-e^{-\phi_0/2}s/8)\Gamma(-e^{-\phi_0/2}t/8)\Gamma(-e^{-\phi_0/2}u/8)}{\Gamma(1+e^{-\phi_0/2}s/8)\Gamma(1+e^{-\phi_0/2}t/8)\Gamma(1+e^{-\phi_0/2}u/8)}&\sim &-\frac{2^{9}e^{3\phi_0/2}}{stu}-2\zeta(3)+\cdots\label{expa}
\eea
where the expansion parameter is $ \alpha' $ and $ \ph_0 $ is the constant dilaton background. The right-hand side in the above expression gives the low-energy limit of the amplitude. The first term in that side just corresponds to amplitudes of exchanging two RR-two NSNS massless fields in $ s $, $ t $ and $ u $ channels as well as contact terms in type II supergravity \cite{Sannan:1986tz}. The second term contains a Riemann zeta function $ \z(3) $, which is irrational, and contributes as a stringy correction to the supergravity. In this way, it is basically possible to derive higher-derivative corrections from string scattering amplitudes.

By comparing the amplitude (\ref{ampf}) with the leading term of string-theory amplitude, one finds the following relation between the field-theory amplitude and the string-theory kinematic factor
\bea
{\cal K}&=&-2^{-9} e^{-3\phi_0/2}\lp tu{\cal A}_s+st {\cal A}_u+su{\cal A}_t+sut{\cal A}_c\rp.
\eea
Multiplying this factor by $ -2\z(3) $ and transforming it to the spacetime, one then finds the couplings of two RR-two NSNS fields at order $ \a'^3 $.

\section{$(\pa F^{(n)})^2 R^2$ couplings}

After explaining our strategy in the previous section, we are now in a position to find various couplings. For the first one, we are going to find the couplings containing two RR $ n $-form field strengths with $ n=1,2,3,4,5 $ and two Riemann curvatures. To this end, we consider the elastic scattering process of two RR fields into two gravitons. When the two RR forms have the same rank, the supergravity actions (\ref{IIA}) and (\ref{IIB}) dictate that the massless poles in the $ s $ and $ u $ channels and the contact terms are given by the following expressions
\bea 
A_s&=&\lp{\td V}_{F_1^{(n)}F_2^{(n)}h}\rp^{\mu \nu}\lp{\td G}_{h} \rp_{\mu \nu, \lambda\rho}\lp{\td V}_{h h_3 h_4}\rp^{\lambda\rho},\nonumber\\ A_u&=&\lp{\td V}_{F_1^{(n)}h_3C^{(n-1)}}\rp^{\mu_1\cdots \mu_{n-1}}\lp{\td G}_{C^{(n-1)}} \rp_{\mu_1\cdots \mu_{n-1}}{}^{\nu_1\cdots \nu_{n-1}}\lp{\td V}_{C^{(n-1)}F_2^{(n)}h_4}\rp_{\nu_1\cdots \nu_{n-1}},\nonumber\\ A_c&=&{\td V}_{F_1^{(n)}F_2^{(n)} h_3 h_4}. \label{sucFnFnRR}
\eea
The $t$-channel amplitude $ A_t $ can now be obtained from the $ A_u $ by permuting
the particles lines 3 and 4.

For the process we want to calculate, we need the two RRs-one graviton, three gravitons, one RR-one graviton-one RR and two RRs-two gravitons vertex functions. These vertices can be obtained from the supergravity actions (\ref{IIA}) and (\ref{IIB}) by expanding to cubic and quartic powers in $ \k $, respectively. Fortunately, these vertex functions as well as the graviton and RR propagators (in Feynman-like gauge) have been derived previously in the literatures \cite{Sannan:1986tz,Garousi:1996ad,Garousi:1998bj,Bakhtiarizadeh:2013zia,Bakhtiarizadeh:2015exa} up to an overall factor:
\bea
\lp{\td V}_{F_1^{(n)}F_2^{(n)}h}\rp^{\mu \nu}&=&  
\frac{i\k}{ n!}\left(2n\,F_{1}^{(\mu}{}_{\mu_1\cdots\mu_{n-1}}F_2^{\nu)\mu_1\cdots\mu_{n-1}}  -\eta^{\mu\nu}\,F_ {1\mu_1\cdots\mu_{n}}F_2^{\mu_1\cdots\mu_{n}} \right), \label{F1F1h}
\eea
\bea
\lp{\td G}_{h} \rp_{\mu \nu, \lambda\rho}&=&
-\frac{i}{2k^2}\lp\eta_{\mu\lambda}\eta_{\nu\rho}+
\eta_{\mu\rho}\eta_{\nu\lambda}-\frac{1}{4}\eta_{\mu\nu}
\eta_{\lambda\rho}\rp,\label{gpro}
\eea
\bea
\lp{\td V}_{h_1 h_2 h}\rp^{\lambda\rho}&=&-2i\kappa\,\ls 
\left(\frac{}{}\frac{3}{2}k_1.
k_2\,\eta^{\lambda\rho}+k_1^{(\lambda}\,k_2^{\rho)}-k^{\lambda}\,k^{\rho}
\right){\rm Tr}( h_1.h_2)\right.
\nonumber\\
&&\qquad\ -k_1.h_2.h_1.k_2\, \eta^{\lambda\rho}+
2\,k_2^{(\lambda}\,{\eta_2}^{\rho)}.h_1.k_2
+2\,{k_1}^{(\lambda}\,{\eta_1}^{\rho)}.h_2.k_1
\nonumber\\
&&\qquad \ +2k_1.{h_2}^{(\lambda}\,{\eta_1}^{\rho)}.k_2
-k_1.k_2\,\left( h_1^{\lambda}.h_2^{\rho}+ h_2^{\lambda}.h_1^\rho\right)
\nonumber\\
&&\qquad\left.\ -k_1.h_2.k_1\, h_1^{\lambda\rho}
-k_2.h_1.k_2\, h_2^{\lambda\rho} \vphantom{\frac{3}{2}}\rs,\label{h1h2h}
\eea
\bea
\lp{\td V}_{F_1^{(n)}h_2C^{(n-1)}}\rp^{\mu_1\cdots \mu_{n-1}}&=& \frac{2i\k}{(n-1)!} F_{1 \m}{}^{ [\mu_1\cdots \mu_{n-1}} h_2^{\m |\n ]} k_{\n},\label{FnhCn-1}
\eea
\bea
\lp{\td G}_{C^{(n)}} \rp_{\mu_1\cdots \mu_{n}}{}^{\nu_1\cdots \nu_{n}}&=&-\frac{in!}{k^2}
\eta_{[\mu_1}{}^{\nu_1}\eta_{\mu_2}{}^{\nu_2}\cdots\eta_{\mu_{n}]}{}^{\nu_{n}},\label{RRpro}
\eea
\bea
{\td V}_{F_1^{(n)}F_2^{(n)} h_3 h_4}&=&\frac{4i\k^2}{n!}\ls F_{1\m_1\cdots\m_{n}}F_{2}^{\m_1\cdots\m_{n}}\lp -\frac{1}{4}h_{3\l \r}h_4^{\l \r}+\frac{1}{8}h_3{}^{\l}{}_{\l}h_4{}^{\r}{}_{\r}\rp \right.\nn\\&&-\frac{1}{2}(n+1) F_{1\m\m_2\cdots\m_{n}}F_{2\n}{}^{\m_2\cdots\m_{n}}h_3{}^{\l}{}_{\l}h_4^{\m \n}\nn\\&&+\frac{1}{2}n(n-1) F_{1\m\l\m_3\cdots\m_{n}}F_{2\n\r}{}^{\m_3\cdots\m_{n}}h_3^{\m \n}h_4^{\l \r}\nn\\&&\left.+ F_{1}{}^{\m\m_2\m_3\cdots\m_{n}}F_{2}{}^{\n}{}_{\m_2\m_3\cdots\m_{n}}h_{3\m}{}^{ \l}h_{4\n\l} \vphantom{\frac{1}{2}}\rs+(3\leftrightarrow 4).
\eea
Our notation is such that $ {\rm Tr}(h_1.h_2)=h_{1}^{\m \n}h_{2\n \m}, k_1.h_2^{\l}=k_{1\r} h_2^{\r\l}$ and $ h_1^{\r}.k_2=h_1^{\r \l}k_{2\l} $. The bracket (parentheses) notation over indices means antisymmetrization (symmetrization) with a factor $ 1/2 $. $ k_{\n} $ in \ref{FnhCn-1} denotes the momentum of internal leg \ie $ k_{\n}=k_{1\n}+k_{2\n} $. 

The next step is to write the amplitude in terms of independent variables. This imposes all symmetries including mono-term symmetries (antisymmetry property of RR potentials and symmetry property of graviton polarizations) as well as multi-term symmetries (the Bianchi identities governing on the RR field strengths and Riemann tensors). In doing so, we have first written the RR field strengths and Riemann tensors in terms of RR potentials and graviton polarizations, respectively. Then, we have manipulated with some of the terms. For instance, momentum conservation $ \Sigma_{i=1}^{4} k_i =0$ as well as the mass-shell and on-shell relations for momenta, $ k_{i}^2 = 0 $ and $ k_i.\e_i=0 $, are imposed. We have also rewritten some terms using the relation $ s+t+u=0 $. Furthermore, we have applied the traceless property of external graviton polarizations.

After doing the above steps to canonicalize the amplitude, the final result is simplified in terms of some Mandelstam variables, momenta, two RR potentials and two graviton polarizations.
When $ n=1 $, the result is simplified as:
\bea
&& \frac{1}{2} \lp s^2 u^2 \ h_{3}^{\alpha \beta} h_{4\alpha \beta} + 2 s u^3 \ h_{3}^{\alpha \beta} h_{4\alpha \beta} +
u^4 \ h_{3}^{\alpha \beta} h_{4\alpha \beta} + 16 s u^2 \ h_{3\beta}{}^{\gamma} h_{4\alpha \gamma} k_{1}^{\alpha} k_{2}^{\beta} \right. \nonumber \\ 
&&+ 16 u^3 \ h_{3\beta}{}^{\gamma} h_{4\alpha \gamma} k_{1}^{\alpha} 
k_{2}^{\beta} - 16 s^2 u \ h_{3\alpha}{}^{\gamma} h_{4\beta \gamma} k_{1}^{\alpha} k_{2}^{\beta}  - 32 s u^2 \ h_{3\alpha}{}^{\gamma} h_{4\beta \gamma} k_{1}^{\alpha} k_{2}^{\beta} \nonumber \\ 
&&- 16 u^3 \ h_{3\alpha}{}^{\gamma} h_{4\beta \gamma} k_{1}^{\alpha} k_{2}^{\beta}  + 64 u^2 \ h_{3\gamma \delta} h_{4\alpha \beta} k_{1}^{\alpha} k_{1}^{\beta} k_{2}^{\gamma}
k_{2}^{\delta} - 128 s u \ h_{3 \alpha \gamma} h_{4\beta \delta}
k_{1}^{\alpha} k_{1}^{\beta} k_{2}^{\gamma} k_{2}^{\delta} \nonumber \\ 
&& - 128 u^2 \ 
h_{3\alpha \gamma} h_{4\beta \delta} k_{1}^{\alpha} k_{1}^{\beta}
k_{2}^{\gamma} k_{2}^{\delta}  + 64 s^2 \ h_{3\alpha \beta} h_{4\gamma \delta} 
k_{1}^{\alpha} k_{1}^{\beta} k_{2}^{\gamma} k_{2}^{\delta} + 128 s u  \
h_{3\alpha \beta} h_{4\gamma \delta} k_{1}^{\alpha} k_{1}^{\beta} 
k_{2}^{\gamma} k_{2}^{\delta} \nonumber \\ 
&& + \left. 64 u^2 \ h_{3\alpha \beta} h_{4\gamma \delta}
k_{1}^{\alpha} k_{1}^{\beta} k_{2}^{\gamma} k_{2}^{\delta}\rp, \label{simplified}
\eea  
where the RR polarizations have been written in terms of momenta. To rewrite the amplitude \ref{simplified} in terms of RR field strengths and Riemann tensors, we consider all possible contractions between two RR $ 1 $-form field strengths with a partial derivative acting on each one and two Riemann tensors. Then, we make an ansatz of these all contractions by multiplying them by unknown constant coefficients:
\bea 
&&  C_1  F^{a,b} F^{c,d} R_{ac}{}^{ef} R_{bdef}+ C_2  F^{a,b} F^{c,d} R_{ac}{}^{ef} R_{bedf} + C_3  F^{a,b} F^{c,d} R_{a}{}^{e}{}_{c}{}^{f} R_{bedf} \nonumber \\&&+ C_4  F^{a,b} F^{c,d} R_{a}{}^{e}{}_{c}{}^{f} R_{bfde} + C_5 F_{a}{}^{,c} F^{a,b} R_{b}{}^{def} R_{cdef} + C_6  F^{a,b} F^{c,d} R_{a}{}^{e}{}_{b}{}^{f} R_{cedf} \nonumber \\&&+ C_7 F_{a}{}^{,c} F^{a,b} R_{b}{}^{def} R_{cedf} +C_8 F_{a,b} F^{a,b} R_{cdef} R^{cdef} + C_9  F_{a,b} F^{a,b} R_{cedf} R^{cdef}.\label{allcontract}
\eea

By comparing the expression \ref{simplified} with the result obtained from writing \ref{allcontract} in terms of independent variables with the same steps as mentioned above, one obtains some algebraic equations among unknown constant coefficients, that are:
\bea
&& \{2 C_{5} + C_{6} + C_{7} + 8 C_{8} + 4 C_{9} = 0, 
2 C_{5} + 2 C_{6} + C_{7} = 0, \\&&-512 + 2 C_{1} + C_{2} + 
C_{3} + 2 C_{6} = 0, 
2 C_{1} + C_{2} - C_{4} + 2 C_{5} + 3 C_{6} + C_{7} = 0, \nn\\&&
2 C_{1} + C_{2} - C_{4} + C_{6} = 0, 
2 C_{1} + C_{2} - C_{4} + 2 C_{5} + C_{6} + C_{7} + 16 C_{8} + 
8 C_{9} = 0, 
\nn\\&& 512 + 4 C_{1} + 2 C_{2} - C_{3} - 3 C_{4} + C_{6} = 0, 2 C_{1} + C_{2} - C_{4} + 8 C_{8} + 4 C_{9} = 0, \nn\\&&-1024 + 
2 C_{1} + C_{2} + 2 C_{3} + C_{4} + 2 C_{5} + 5 C_{6} + 
C_{7} = 0, -512 + C_{3} + C_{4} + C_{6} = 0, 
\nn\\&& -1024 + 
4 C_{1} + 2 C_{2} + 2 C_{3} - 2 C_{5} + C_{6} - C_{7} + 
8 C_{8} + 4 C_{9} = 0,  \nn\\&&
6 C_{1} + 3 C_{2} - 3 C_{4} - 2 C_{5} + C_{6} - 
C_{7} = 0,  -512 + C_{3} + C_{4} + 2 C_{5} + 3 C_{6} + 
C_{7} = 0,\nn\\&&-4096 + 16 C_{1} + 8 C_{2} + 8 C_{3} - 8 C_{5} + 
8 C_{6} - 4 C_{7} = 0,  \nn\\&&
512 + 2 C_{1} + C_{2} - C_{3} - 2 C_{4} = 0,
2048 + 2 C_{1} + C_{2} - 4 C_{3} - 5 C_{4} + 2 C_{5} - C_{6} + 
C_{7} = 0\}\nn
\eea
After solving the above eighteen equations, simultaneously, one finds the unknown constant coefficient $ C_{3} $ to be $ 512 $. The other coefficients are free parameters which can be set to zero. 

Having had the unknown coefficients and substituting them into the expression \ref{allcontract}, we can obtain the couplings between two RR 1-form field strengths and two Riemann tensors\footnote{The calculations in this study have been carried out by the Mathematica package xAct\cite{xact,Nutma:2013zea}.}. The effective action, which reproduces the corresponding amplitude, takes the following form in the string frame
\bea
S_{(\pa F^{(1)})^2 R^2} = \frac{\gamma}{3.2^{4}\kappa^2} \int d^{10}x \ e^{2\phi_{0}} \sqrt{-G} \  F^{a,b} F^{c,d} R_{a}{}^{e}{}_{c}{}^{f} R_{bedf},\label{F1F1RR}
\eea
where $ \g=\a'^3 \z(3)/2^5 $, and comma refers to a partial derivative with respect to the index afterwards. Note that, to acquire the standard sphere-level dilaton factor $ e^{-2 \phi_{0}} $ in the string frame, it is convenient to normalize the RR potentials and graviton polarizations with a factor $ e^{\phi_0} $. The normalization of RR fields in the above action are consistent with the supergravities (\ref{IIA}) and (\ref{IIB}). Moreover, the couplings of two RR 2-form field strengths and two Riemann curvatures appear in the following effective action
\bea
S_{(\pa F^{(2)})^2 R^2} &=& \frac{\gamma}{3.2^{5}\kappa^2} \int d^{10}x \ e^{2\phi_{0}} \sqrt{-G} \  \left(2 F_{a}{}^{d,e} F^{ab,c} R_{b}{}^{f}{}_{d}{}^{g} R
_{cfeg}  +  
F^{ab,c} F_{c}{}^{d,e} R_{ad}{}^{fg} R_{befg} \right.\nn\\&&\left.+ 2  F^{ab,c}
F^{de,f} R
_{acf}{}^{g} R_{bdeg} + 2  F^{ab,c} F^{de,f} R_{afd}{}^{g} R
_{bgce}\right). \label{F2F2RR}
\eea

Hence, there is no unique way of writing down the expressions (\ref{F1F1RR}) and (\ref{F2F2RR}) because of different presentations of tensor polynomials using the symmetries of the individual tensors, but we believe the above forms are the most economical one. The number of terms has been reduced as much as possible using the algorithm introduced in \cite{Bakhtiarizadeh:2015exa}. We also find the couplings between two RR 3-form field strengths and two Riemann curvatures as
\bea
S_{(\pa F^{(3)})^2 R^2} &=& \frac{\gamma}{3^2.2^{7}\kappa^2} \int d^{10}x \ e^{2\phi_{0}} \sqrt{-G} \ \left(8  F_{abc}{}^{,e} F^{abc,d} R_{d}{}^{fgh} R
_{efgh}\right.\nn\\&&- 3 F_{abd,c} F^{abc,d} R_{efgh} R^{efgh} + 12  
F_{ab}{}^{e,f} F^{abc,d} R
_{cf}{}^{gh} R_{dehg} \nn\\&&- 12 
F_{ab}{}^{e,f} F^{abc,d} R
_{ce}{}^{gh} R_{dfgh}  + 24 
F_{ab}{}^{e,f} F^{abc,d} R
_{cd}{}^{gh} R_{efgh} \nn\\&&- 12 
F_{ab}{}^{e,f} F^{abc,d} R
_{cd}{}^{gh} R_{efhg} + 24 
F_{ad}{}^{e}{}_{,b} F^{abc,d} R
_{c}{}^{fgh} R_{efgh} \nn\\&&- 24 F_{a}{}^{ef,g} F^{abc,d} R_{bde}{}^{h} R_{cfhg}+ 48 F_{a}{}^{ef,g} F^{abc,d} R
_{bce}{}^{h} R_{dfgh}\nn\\&&+ 24 
F_{a}{}^{ef,g} F^{abc,d} R_{bce}{}^{h} R_{dfhg} - 24 F_{a}{}^{ef,g} F^{abc,d} R_{bcd}{}^{h} R_{efgh} \nn\\&&\left.- 12  F^{abc,d} F_{d}{}^{ef}{}_{,a} R_{be}{}^{gh} R_{cfhg}+ 12  F^{abc,d} 
F_{d}{}^{ef}{}_{,a} R_{bc}{}^{gh} R_{efgh}\right).
\eea

Furthermore, our calculations show that the couplings which include two RR 4-form field strengths and two Riemann curvatures have the following form
\bea
S_{(\pa F^{(4)})^2 R^2} &=& -\frac{\gamma}{3^2.2^{6}\kappa^2} \int d^{10}x \ e^{2\phi_{0}} \sqrt{-G} \  \left(3 
F_{abef}{}^{,i} F_{cdgi,h} R^{abcd} R^{efgh} \right. \nn\\&&- 9 F_{abe}{}^{i}{}_{,f} F_{cdgi,h} R^{abcd} R^{efgh}- 9 F_{abe}{}^{i}{}_{,g} F_{cdhi,f} R^{abcd} R^{efgh} \nn\\&&+ 6 F_{be}{}^{gh,i} F_{dfgi,h} R_{a}{}^{e}{}_{c}{}^{f} R^{abcd} + 3
F_{ce}{}^{gh,i} F_{dfgi,h} R_{ab}{}^{ef} R^{abcd} \nn\\&&+ 6 F_{be}{}^{hi}{}_{,c} F_{dfhi,g} R_{a}{}^{efg} R^{abcd} - 6 F_{cf}{}^{hi}{}_{,b} F_{dghi,e} R_{a}{}^{efg} R^{abcd} \nn\\&&- 2 
F_{b}{}^{ghi}{}_{,e} F_{dghi,f} R_{a}{}^{e}{}_{c}{}^{f} R^{abcd}+ 3 
F_{ce}{}^{gh,i} F_{dghi,f} R_{ab}{}^{ef} R^{abcd}\nn\\&&- 24 F_{bce}{}^{h,i} F_{fghi,d} R_{a}{}^{efg} R^{abcd} - 6 F_{bc}{}^{hi}{}_{,e} F_{fghi,d} R_{a}{}^{efg} R^{abcd} \nn\\&&\left.+ 9 F_{be}{}^{hi}{}_{,c} F_{fghi,d}
R_{a}{}^{efg} R^{abcd} \right).
\eea

We are now going to find the couplings containing two RR 5-form field strengths and two Riemann tensors. The amplitude in this case is somehow different, because of the presence of the 5-form field strength $ F^{(5)} $ which is self-dual. This condition should be imposed manually as an extra condition in physical quantities like the equations of motion as well as S-matrix.

We expect that imposing the self-duality condition $ F^{(5)}\to \left(F^{(5)}+\star F^{(5)} \right)/2 $, which is equivalent to (\ref{selfdual}), in the couplings with structure $ (\pa F^{(5)})^2 R^2 $ not only gives the correct overall factor but also results into another coupling with structure $ \e_{10} (\pa F^{(5)})^2 R^2 $. This coupling is not calculated here, because the number of indices is too large that make it difficult to calculate all possible contractions for this structure. However, the result for the couplings between two RR 5-form field strengths and two Riemann curvatures is 
\bea
S_{(\pa F^{(5)})^2 R^2} &=& \frac{\gamma}{3^2.2^{8}\kappa^2} \int d^{10}x \ e^{2\phi_{0}} \sqrt{-G} \ \left(18 F_{abe}{}^{ij}{}_{,g} F_{cdfij,h} R^{abcd} R^{efgh} \right.\nn\\&&- 3 F_{abef}{}^{i,j} F_{cdghi,j}
R^{abcd} R^{efgh} - 12 F_{be}{}^{ghi,j} F_{dfghj,i} R_{a}{}^{e}{}_{c}{}^{f} R^{abcd} \nn\\&&- 6 F_{ce}{}^{ghi,j} F_{dfghj,i} R_{ab}{}^{ef} R^{abcd} - 8 F_{be}{}^{hij}{}_{,c} F_{dfhij,g} R_{a}{}^{efg} R^{abcd}\nn\\&&+ 8 F_{cf}{}^{hij}{}_{,b} F_{dghij,e} R_{a}{}^{efg} R^{abcd} + 2 F_{b}{}^{ghij}{}_{,e} F_{dghij,f} R_{a}{}^{e}{}_{c}{}^{f} R^{abcd}\nn\\&&+ 4 F_{ce}{}^{ghi,j} F_{dghij,f} R_{ab}{}^{ef} R^{abcd} - 48 F_{bce}{}^{hi,j} F_{fghij,d} R_{a}{}^{efg} R^{abcd}\nn\\&&\left.+ 8 F_{bc}{}^{hij}{}_{,e} F_{fghij,d} R_{a}{}^{efg} R^{abcd}- 12  F_{be}{}^{hij}{}_{,c} F_{fghij,d} R_{a}{}^{efg} R^{abcd}\right).
\eea
where the product of two Levi-Civita tensors implicitly has been replaced by the generalized Kronecker delta according to:
$
\epsilon^{m_1 \cdots m_{d}}\epsilon_{n_1 \cdots n_{d}}= -\delta_{[n_1}{}^{m_1} \cdots \delta_{n_d]}{}^{m_d}.
$

In the next sections we follow a similar approach to obtain the other couplings. The detail of calculations are omitted there for the sake of brevity. 
\section{$(\pa F^{(n)})^2 (\pa \ph)^2$ couplings}

In this section, the couplings with structure $(\pa F^{(n)})^2 (\pa \ph)^2$ will be calculated. Similar to the previous section, there are also five types of couplings here, \ie with $ n=1,2,3,4,5 $. When we have two RR field strengths, in which the RR forms have the same rank, and two dilatons, the actions (\ref{IIA}) and (\ref{IIB}) indicate that the amplitudes in the $ s $ and $ u $ channels and the contact terms become
\bea 
A_s&=&\lp{\td V}_{F_1^{(n)}F_2^{(n)}h}\rp^{\mu \nu}\lp{\td G}_{h} \rp_{\mu \nu, \lambda\rho}\lp{\td V}_{h \phi_3 \phi_4}\rp^{\lambda\rho},\nonumber\\A_u&=&\lp{\td V}_{F_1^{(n)}\phi_3C^{(n-1)}}\rp^{\mu_1\cdots \mu_{n-1}}\lp{\td G}_{C^{(n-1)}} \rp_{\mu_1\cdots \mu_{n-1}}{}^{\nu_1\cdots \nu_{n-1}}\lp{\td V}_{C^{(n-1)}F_2^{(n)}\phi_4}\rp_{\nu_1\cdots \nu_{n-1}},\nonumber\\ A_c&=&{\td V}_{F_1^{(n)}F_2^{(n)} \phi_3 \phi_4}, \label{sucFnFnDD}
\eea
respectively. Here also the amplitude in the $t$ channel is the same as $A_u$ in which the particle labels of the RR fields are interchanged, \ie $A_t= A_u(3 \leftrightarrow 4)$. The required tree- and four-point interactions are given by
\bea
\lp{\td V}_{\phi_1 \phi_2 h}\rp^{\m \n}&=&-2i\kappa  \lp k_1^{(\m} k_2^{\n)} - \frac{1}{2} k_1 . k_2 \eta^{\m \n}\rp ,\label{DDh}
\eea
\bea
\lp \td V_{F_1^{(n)}\phi_2 C^{(n-1)}} \rp_{\nu_1\cdots\nu_{n-1}}&=& i\kappa\frac{1}{\sqrt{2}(n-1)!}(5-n)\,F_{1\m\nu_1\cdots\nu_{n-1}}k^{\m}, \label{FDC}
\eea
\bea
{\td V}_{F_1^{(n)}F_2^{(n)}\phi_3\phi_4}&=&-2i\kappa^2 \frac{1}{n!}\left(\frac{5-n}{2}\right)^2 F_{1 \m_1 \cdots \m_{n}}F_{2}^{\m_1 \cdots \m_{n}}.\label{FFDD}
\eea
The graviton and RR propagators are given by equations \ref{gpro} and \ref{RRpro}, respectively. Substituting these expressions into the amplitude \ref{sucFnFnDD}, one finds the couplings consisting of two RR 1-form field strengths and two dilatons as follow
\bea
S_{(\pa F^{(1)})^2 (\pa \ph)^2} = \frac{\gamma}{3.2^{5}\kappa^2} \int d^{10}x \ e^{2\phi_{0}} \sqrt{-G} \  \left(2 F^{a,b} F^{c,d} \phi_{a,b}
\phi_{c,d} -  F_{a,b} F^{a,b}
\phi_{c,d} \phi^{c,d}\right).
\eea

One can also write down the couplings between two RR 2-form field strengths and two dilatons as
\bea
S_{(\pa F^{(2)})^2 (\pa \ph)^2} &=& \frac{\gamma}{3.2^{8}\kappa^2} \int d^{10}x \ e^{2\phi_{0}} \sqrt{-G} \  \left(16 F_{a}{}^{d,e} F^{ab,c} 
\phi_{b,d} \phi_{c,e} \right.\nn\\&&\left.+ 18 
F_{a}{}^{d}{}_{,c} F^{ab,c} 
\phi_{b}{}^{,e} \phi_{d,e} - 9 F_{ac,b}
F^{ab,c} \phi_{d,e} \phi^{d,e}\right).
\eea

In the same way, the couplings of two RR 3-form field strengths and two dilatons can be written as
\bea
S_{(\pa F^{(3)})^2 (\pa \ph)^2} &=& -\frac{\gamma}{3^2.2^{6}\kappa^2} \int d^{10}x \ e^{2\phi_{0}} \sqrt{-G} \  \left(F_{def,c} 
F^{def}{}_{,b} \phi_{a}{}^{,c} \phi^{a,b} \right.\nn\\&&\left.- 3 F_{a}{}^{e,f}{}_{,b} F_{cef,d} 
\phi^{a,b} \phi^{c,d} - 3 F_{a}{}^{ef}{}_{,c} 
F_{def,b} \phi^{a,b} \phi^{c,d}\right).
\eea

Furthermore, the couplings of two RR 4-form field strengths and two dilatons can be introduced as
\bea
S_{(\pa F^{(4)})^2 (\pa \ph)^2} &=& -\frac{\gamma}{3^2.2^{9}\kappa^2} \int d^{10}x \ e^{2\phi_{0}} \sqrt{-G} \  \left(F_{cdeg,f} F^{cdef,g} 
\phi_{a,b} \phi^{a,b} \right.\nn\\&&\left.- 2 F_{b}{}^{def,g} 
F_{cdef,g} \phi_{a}{}^{,c} \phi^{a,b} - 16 
F_{a}{}^{efg}{}_{,c} F_{befg,d} 
\phi^{a,b} \phi^{c,d}\right).
\eea

Finally, for $ n=5 $ case only the contribution of $ s $ channel exists and is given by the first line in (\ref{sucFnFnDD}). The amplitudes in the $ t $ and $ u $ channels and the contact terms here are zero. Once more, using the self-duality condition, one can easily find the following coupling for two RR 5-form field strengths and two dilatons with correct overall factor
\bea
S_{(\pa F^{(5)})^2 (\pa \ph)^2} = \frac{\gamma}{3^2.2^{7}\kappa^2} \int d^{10}x \ e^{2\phi_{0}} \sqrt{-G} \ F_{a}{}^{efgh}{}_{,c} 
F_{befgh,d} \phi^{a,b} \phi^{c,d}.
\eea
Note that, imposing the self-duality condition also leads to a new coupling of the form $\e_{10}(\pa F^{(5)})^2 (\pa \ph)^2$, but we ignore it for the reasons already mentioned. We use the algorithm introduced in \cite{Bakhtiarizadeh:2015exa} to reduce the tensor polynomials and rewrite the couplings of this section in the minimal-term form.
\section{$(\pa F^{(n)})^2 (\pa \ph)R$ couplings}

Let us now consider the couplings with structure $ (\pa F^{(n)})^2 (\pa \ph)R $. At the first glance, it seems that there are five types of couplings in this section, \ie with $ n=1,2,3,4,5 $, but the contribution of $ \pa F^{(1)} \pa F^{(1)} \pa \phi R $ and $ \pa F^{(5)} \pa F^{(5)} \pa \phi R $ vanishes as can be seen from the action (\ref{IIB}). When the two RR forms have the same rank, from the type II supergravity actions, it is concluded that the amplitudes in the $ s $ and $ u $ channels and the contact terms are given by the following expressions
\bea 
A_s&=&{\td V}_{F_1^{(n)}F_2^{(n)}\phi}{\td G}_{\phi} {\td V}_{\phi \phi_3 h_4},\nonumber\\A_u&=&\lp{\td V}_{F_1^{(n)}\phi_3C^{(n-1)}}\rp^{\mu_1\cdots \mu_{n-1}}\lp{\td G}_{C^{(n-1)}} \rp_{\mu_1\cdots \mu_{n-1}}{}^{\nu_1\cdots \nu_{n-1}}\lp{\td V}_{C^{(n-1)}F_2^{(n)}h_4}\rp_{\nu_1\cdots \nu_{n-1}},\nonumber\\ A_c&=&{\td V}_{F_1^{(n)}F_2^{(n)} \phi_3 h_4}, \label{sucFnFnRD}
\eea
respectively. The amplitude in the $t$ channel is the same as $A_u$ in which the particle labels of the RR fields are interchanged, \ie $A_t= A_u(3 \leftrightarrow 4)$. Here, the dilaton propagator and vertex operators which are needed in the calculation of amplitude, are:
\bea
{\td V}_{F_1^{(n)}F_2^{(n)}\phi} &=& -i\kappa\frac{5-n}{\sqrt{2}n!}\,F_{1 \m_1 \cdots \m_{n}}F_{2}^{\m_1 \cdots \m_{n}}, 
\eea 
\bea
{\td G}_{\phi} &=&-\frac{i}{k^2},\label{Dpro}
\eea
\bea
{\td V}_{h_1 \phi_2 \phi}&=& -2i\k k_2.h_1.k, \label{DhD}
\eea
\bea
\lp{\td V}_{F_1^{(n)}\phi_2 C^{(n-1)}}\rp^{\mu_1\cdots \mu_{n-1}}&=&-i\kappa\frac{5-n}{\sqrt{2}(n-1)!}\,F_{1}{}^{\lambda\mu_1\cdots\mu_{n-1}}k_{\lambda},
\eea
\bea
{\td V}_{F_1^{(n)}F_2^{(n)} \phi_3 h_4}&=&\frac{i\k^2(5-n)}{\sqrt{2}n!} \lp F_{1\m_1 \cdots \m_{n}}F_2^{\m_1 \cdots \m_{n}} h_{4\m}{}^{\m} -2n F_{1 \m \m_2 \cdots \m_{n}}F_{2 \n}{}^{\m_2 \cdots \m_{n}} h_4^{\m \n} \rp.\label{FFDh}
\eea
The RR propagator is given by equation \ref{RRpro}, as well. After doing some algebra, we have
\bea
S_{(\pa F^{(2)})^2 (\pa \ph)R} &=& \frac{\gamma}{2^{\frac{11}{2}}\kappa^2} \int d^{10}x \ e^{2\phi_{0}} \sqrt{-G} \ \left(2 
F_{a}{}^{d,e} F^{ab,c} 
\phi_{b}{}^{,f} R_{cdef} \right.\nn\\&&\left.- 2  
F^{ab,c} F_{c}{}^{d,e} 
\phi_{e}{}^{,f} R_{adbf} -    
F_{a}{}^{d}{}_{,c} F^{ab,c} \phi^{e,f} R_{bedf}\right).
\eea

Doing the same steps as above, one finds the couplings with structure $ (\pa F^{(2)})^2 (\pa \ph)R $ in the type IIA theory. In the string frame, they are
\bea
S_{(\pa F^{(3)})^2 (\pa \ph)R} &=& -\frac{\gamma}{3.2^{\frac{9}{2}}\kappa^2} \int d^{10}x \ e^{2\phi_{0}} \sqrt{-G} \ \left(2  
F_{a}{}^{cd,e} F_{ce}{}^{f,g} 
\phi^{a,b} R_{bfdg}\right. \\&&\left.-   
F_{a}{}^{cd}{}_{,b} F_{c}{}^{ef,g}
\phi^{a,b} R_{dgef} - 2 F_{a}{}^{cd,e} F_{c}{}^{fg}{}_{,b}
\phi^{a,b} R_{dfeg} + 2   
F_{c}{}^{fg}{}_{,b} F^{cde}{}_{,a} 
\phi^{a,b}\nn\right).
\eea

Finally, we have found the following result for the couplings containing two RR 4-form field strengths, one dilaton and one Riemann curvature in the string frame
\bea
S_{(\pa F^{(4)})^2 (\pa \ph)R} &=& \frac{\gamma}{3^2.2^{\frac{13}{2}}\kappa^2} \int d^{10}x \ e^{2\phi_{0}} \sqrt{-G} \ \left(2
F_{b}{}^{fgh}{}_{,d} F_{cfgh,e} 
\phi^{a,b} R_{a}{}^{cde}\right.\nn\\&&\left.- 3   
F_{bc}{}^{fg,h} F_{defg,h} 
\phi^{a,b} R_{a}{}^{cde} +    F_{c}{}^{efg,h} F_{defg,h} 
\phi^{a,b} R_{a}{}^{c}{}_{b}{}^{d}\right).
\eea 
Here also the algorithm introduced in \cite{Bakhtiarizadeh:2015exa} has been used to rewrite the couplings of this section in their minimal-term
form.
\section{$(\pa F^{(n)})^2 (\pa H)^2$ couplings}

Now, we consider the couplings with structure $(\pa F^{(n)})^2 (\pa H)^2$. There are also five types of couplings in this section, \ie with $ n=1,2,3,4,5 $. When $ n=1,2,3 $, it can be seen from the actions (\ref{IIA}) and (\ref{IIB}) that the $s$- and $u$-channel amplitudes as well as the contact terms are given by
\bea 
A_s&=&\lp{\td V}_{F_1^{(n)}F_2^{(n)}h}\rp^{\mu \nu}\lp{\td G}_{h} \rp_{\mu \nu, \lambda\rho}\lp{\td V}_{h b_3 b_4}\rp^{\lambda\rho}+{\td V}_{F_1^{(n)}F_2^{(n)}\phi}{\td G}_{\phi} {\td V}_{\phi b_3 b_4},\nonumber\\A_u&=&\lp{\td V}_{F_1^{(n)}b_3C^{(n+1)}}\rp^{\mu_1\cdots \mu_{n+1}}\lp{\td G}_{C^{(n+1)}} \rp_{\mu_1\cdots \mu_{n+1}}{}^{\nu_1\cdots \nu_{n+1}}\lp{\td V}_{C^{(n+1)}F_2^{(n)}b_4}\rp_{\nu_1\cdots \nu_{n+1}},\nonumber\\ A_c&=&{\td V}_{F_1^{(n)}F_2^{(n)} b_3 b_4}, \label{sucFnFnHH123}
\eea
respectively. The $t$-channel amplitude $ A_t $ can now be obtained from the $ A_u $ by permuting the external B-fields lines 3 and 4. The vertices, that have not been introduced in the previous sections, are: 
\bea
\lp \td V_{b_1b_2h}\rp^{\m\n}&=&-2i\kappa\,\ls
\frac{1}{2}\lp k_1.
k_2\,\eta^{\m\n}-k_1^{\m}\,k_2^{\n}-k_1^{\n}\,k_2^{\m}
\right){\rm Tr}( b_1. b_2)\right.
\nonumber\\
&&\quad\qquad-k_1. b_2 . b_1 . k_2\, \eta^{\m\n}+
2\,k_1^{(\m}\,{ b_2}^{\n)}. b_1. k_2
+2\,{k_2}^{(\m}\,{ b_1}^{\n)}. b_2. k_1
\nonumber\\
&&\quad\qquad\left.+2k_1.{ b_2}^{(\m}\,{ b_1}^{\n)}. k_2
-k_1.k_2\,( b_1^{\m}.b_2^{\n}+ b_2^{\m}.b_1^\n)\vphantom{\frac{1}{2}}\rs, \label{bbh}
\eea
\bea
{\td V}_{b_1 b_2\phi}&=&- \sqrt{2} i\k \ls 2k_1. b_2. b_1. k_2
-k_1. k_2{\rm Tr}( b_1. b_2)\rs,\label{bbD}
\eea
\bea
\lp \td V_{F_1^{(n)} b_2  C^{(n+1)}}\rp_{\nu_1\cdots\nu_{n+1}}&=& -2i\kappa\frac{1}{(n+1)!}k^{\m} b_{2\m[\nu_1}F_{1\nu_2\cdots\nu_{n+1}]},\label{FnbCn+1}
\eea
\bea
\td V_{F_1^{(n)}F_2^{(n)}b_3b_4}&=&-\frac{2i\kappa^2(n+2)(n+1)}{n!}b_{3[\mu\nu}F_{1\mu_1\cdots\mu_{n}]}b_4^{\mu\nu}F_2^{\mu_1\cdots\mu_{n}}+(3\leftrightarrow 4).
\eea
Putting theses contributions together, one finds the following couplings between two RR 1-form field strengths and two B-field strengths
\bea
S_{(\pa F^{(1)})^2 (\pa H)^2} &=& \frac{\gamma}{3^2.2^{6}\kappa^2} \int d^{10}x \ e^{2\phi_{0}} \sqrt{-G} \  \left(12 F^{a,b} F^{c,d} 
H_{a}{}^{ef}{}_{,c} H_{bef,d} \right. \nn\\&&\left.+ 6 
F_{a}{}^{,c} F^{a,b} 
H_{b}{}^{de,f} H_{cde,f} -  
F_{a,b} F^{a,b} H_{cde,f} 
H^{cde,f}\right), 
\eea
which is written in its minimal-term form. In a similar way, the couplings between two RR 2-form field strengths and two B-field strengths are appeared in the action
\bea
S_{(\pa F^{(2)})^2 (\pa H)^2} &=& -\frac{\gamma}{3^2.2^{5}\kappa^2} \int d^{10}x \ e^{2\phi_{0}} \sqrt{-G} \  \left(3 F^{ab,c} F^{de}{}_{,c} 
H_{ad}{}^{f,g} H_{beg,f} \right. \nn\\&&+ 6 
F^{ab,c} F^{de,f} 
H_{ac}{}^{g}{}_{,b} H_{dfg,e} - 3 
F_{a}{}^{d,e} F^{ab,c} 
H_{b}{}^{fg}{}_{,c} H_{dfg,e} \nn\\&&+ 3 
F^{ab,c} F_{c}{}^{d,e} 
H_{d}{}^{fg}{}_{,a} H_{efg,b} + 3 
F^{ab,c} F^{de,f} 
H_{ab}{}^{g}{}_{,d} H_{efg,c} \nn\\&&- 3 
F_{a}{}^{d,e} F^{ab,c} 
H_{b}{}^{fg}{}_{,d} H_{efg,c} - 3 
F_{a}{}^{d,e} F^{ab,c} 
H_{b}{}^{fg}{}_{,c} H_{efg,d} \nn\\&&+ 3 
F_{a}{}^{d,e} F^{ab,c} 
H_{c}{}^{fg}{}_{,b} H_{efg,d} + 2 
F_{ac}{}^{,d} F^{ab,c}
H_{efg,d} H^{efg}{}_{,b} \nn\\&&\left.-  
F_{a}{}^{d}{}_{,c} F^{ab,c} 
H_{efg,d} H^{efg}{}_{,b}\right).
\eea

It is also straightforward to find the couplings between two RR 3-form field strengths and two B-field strengths as

\bea
S_{(\pa F^{(3)})^2 (\pa H)^2} &=& \frac{\gamma}{3^3.2^{6}\kappa^2} \int d^{10}x \ e^{2\phi_{0}} \sqrt{-G} \ \left(18 F^{abc,d} F^{efg,h} 
H_{abd,h} H_{cef,g} \right.\nn\\&&+ 27 
F_{a}{}^{ef,g} F^{abc,d} 
H_{bcg}{}^{,h} H_{def,h}  + 36 
F_{ab}{}^{e,f} F^{abc,d} 
H_{cf}{}^{g,h} H_{deg,h} \nn\\&&- 36 
F_{a}{}^{ef,g} F^{abc,d} 
H_{bg}{}^{h}{}_{,c} H_{deh,f}  - 9 F^{abc,d} F^{efg,h} H_{abh,e} 
H_{dfg,c} \nn\\&&- 36 F_{a}{}^{ef,g} 
F^{abc,d} H_{bc}{}^{h}{}_{,e} 
H_{dfg,h}  + 36 F_{a}{}^{ef,g} 
F^{abc,d} H_{bg}{}^{h}{}_{,e} 
H_{dfh,c} \nn\\&&+ 72 F^{abc,d} 
F^{efg,h} H_{abe,c} 
H_{dfh,g}  - 36 F_{a}{}^{ef,g} 
F^{abc,d} H_{bc}{}^{h}{}_{,e} 
H_{dfh,g} \nn\\&&+ 54 F_{ab}{}^{e,f} 
F^{abc,d} H_{cf}{}^{g,h} 
H_{dgh,e}  - 9 F_{ab}{}^{e,f} 
F^{abcd} H_{c}{}^{gh}{}_{,f} 
H_{dgh,e} \nn\\&&+ 216 F_{a}{}^{ef,g} 
F^{abc,d} H_{be}{}^{h}{}_{,c} 
H_{dgh,f}  + 72 F_{ab}{}^{e,f} 
F^{abc,d} H_{ce}{}^{g,h} 
H_{dgh,f} \nn\\&&+ 18 F_{ab}{}^{e,f} 
F^{abc,d} H_{c}{}^{gh}{}_{,e} 
H_{dgh,f}  -  F^{abc,d} F^{efg,h} 
H_{abc,d} H_{efg,h} \nn\\&&+ 27 
F_{a}{}^{ef,g} F^{abc,d} 
H_{bcd}{}^{,h} H_{efg,h}  - 18 
F^{abc,d} F_{d}{}^{ef}{}_{,a} 
H_{bc}{}^{g,h} H_{efg,h} \nn\\&&+ 108 
F_{ad}{}^{e,f} F^{abc,d} 
H_{b}{}^{gh}{}_{,c} H_{efg,h}  - 72 F_{ab}{}^{e,f} F^{abc,d} 
H_{cd}{}^{g,h} H_{efg,h} \nn\\&&- 9 
F_{ab}{}^{e}{}_{,d} F^{abc,d} 
H_{c}{}^{fg,h} H_{efg,h} + 36 
F_{ad}{}^{e}{}_{,b} F^{abc,d} 
H_{c}{}^{fg,h} H_{efg,h} \nn\\&& \left.+ 9 
F_{ab}{}^{e,f} F^{abc,d} 
H_{d}{}^{gh}{}_{,c} H_{fgh,e}\right).
\eea

For $ n=4 $, the contribution of $ u $ channel in the amplitude is replaced by
\bea 
A_u&=&\lp{\td V}_{F_1^{(n)}b_3C^{(n-3)}}\rp^{\mu_1\cdots \mu_{n-3}}\lp{\td G}_{C^{(n-3)}} \rp_{\mu_1\cdots \mu_{n-3}}{}^{\nu_1\cdots \nu_{n-3}}\lp{\td V}_{C^{(n-3)}F_2^{(n)}b_4}\rp_{\nu_1\cdots \nu_{n-3}}\nonumber\\&&+\lp{\td V}_{F_1^{(n)}b_3C^{(n+1)}}\rp^{\mu_1\cdots \mu_{n+1}}\lp{\td G}_{C^{(n+1)}} \rp_{\mu_1\cdots \mu_{n+1}}{}^{\nu_1\cdots \nu_{n+1}}\lp{\td V}_{C^{(n+1)}F_2^{(n)}b_4}\rp_{\nu_1\cdots \nu_{n+1}}. \label{uFnFnHH45}
\eea
where
\bea
\lp \td V_{F_1^{(n)} b_2  C^{(n-3)}}\rp_{\nu_1\cdots\nu_{n-3}}&=&-i\kappa\frac{1}{(n-3)!}F_{1\m \n \r \nu_1\cdots\nu_{n-3}}b_2^{\m \n}k^{\r}.\label{FnbCn-3}
\eea
Inserting this contribution into the amplitude, one finds the following couplings between two RR 4-form field strengths and two B-field strengths  
\bea
S_{(\pa F^{(4)})^2 (\pa H)^2} &=& -\frac{\gamma}{3^2.2^{8}\kappa^2} \int d^{10}x \ e^{2\phi_{0}} \sqrt{-G} \  \left(F_{fghi,e} 
F^{fghi}{}_{,d} H_{abc}{}^{,e} 
H^{abc,d} \right.\nn\\&&- 2 F_{efgi,h} 
F^{efgh,i} H_{abd,c} H^{abc,d} - 
8 F_{d}{}^{ghi}{}_{,c} F_{fghi,e} 
H_{ab}{}^{e,f} H^{abc,d} \nn\\&&+ 12 
F_{bc}{}^{gh,i} F_{efgi,h} 
H_{a}{}^{ef}{}_{,d} H^{abc,d} - 8 
F_{b}{}^{ghi}{}_{,c} F_{eghi,f} 
H_{a}{}^{ef}{}_{,d} H^{abc,d} \nn\\&&+ 48 
F_{bd}{}^{hi}{}_{,e} F_{cfhi,g} 
H_{a}{}^{ef,g} H^{abc,d} - 48 F_{bg}{}^{hi}{}_{,c} F_{dehi,f} 
H_{a}{}^{ef,g} H^{abc,d} \nn\\&&- 24 
F_{bc}{}^{hi}{}_{,e} F_{dfhi,g} 
H_{a}{}^{ef,g} H^{abc,d}  - 48 
F_{be}{}^{hi}{}_{,c} F_{dfhi,g}
H_{a}{}^{ef,g} H^{abc,d} \nn\\&&- 48 
F_{be}{}^{hi}{}_{,c} F_{dghi,f} 
H_{a}{}^{ef,g} H^{abc,d}  - 24 
F_{bcd}{}^{h,i} F_{efgi,h} 
H_{a}{}^{ef,g} H^{abc,d} \nn\\&&- 6 
F_{bc}{}^{hi}{}_{,g} F_{efhi,d} 
H_{a}{}^{ef,g} H^{abc,d}  + 24 
F_{bcd}{}^{h,i} F_{efhi,g} 
H_{a}{}^{ef,g} H^{abc,d} \nn\\&& \left.+ 6 
F_{bc}{}^{hi}{}_{,d} F_{efhi,g} 
H_{a}{}^{ef,g} H^{abc,d} - 96 
F_{bcd}{}^{h,i} F_{eghi,f} 
H_{a}{}^{ef,g} H^{abc,d}\right).
\eea

For $ n=5 $, in addition to the $ u $-channel amplitude which is given by (\ref{uFnFnHH45}), the $ s $-channel amplitude is also replaced by
\bea 
A_s&=&\lp{\td V}_{F_1^{(n)}F_2^{(n)}h}\rp^{\mu \nu}\lp{\td G}_{h} \rp_{\mu \nu, \lambda\rho}\lp{\td V}_{h b_3 b_4}\rp^{\lambda\rho}.\label{sFnFnHH45}
\eea
Applying these changes into the amplitude along with imposing the self-duality condition, after some simplifications, one finds the following couplings between two RR $ 5 $-form field strengths and two B-field strengths

\bea
S_{(\pa F^{(5)})^2 (\pa H)^2} &=&- \frac{\gamma}{5.3^2.2^{9}\kappa^2} \int d^{10}x \ e^{2\phi_{0}} \sqrt{-G} \  \left(2 F_{fghij,e} 
F^{fghij}{}_{,d} H_{abc}{}^{,e} 
H^{abc,d} \right.\\&&- 5 F_{efghj,i} 
F^{efghi,j} H_{abd,c} H^{abc,d}  - 
20 F_{d}{}^{ghij}{}_{,c} F_{fghij,e} 
H_{ab}{}^{e,f} H^{abc,d} \nn\\&&+ 60 
F_{bc}{}^{ghi,j} F_{efghj,i} 
H_{a}{}^{ef}{}_{,d} H^{abc,d}  - 20 
F_{b}{}^{ghij}{}_{,c} F_{eghij,f} 
H_{a}{}^{ef}{}_{,d} H^{abc,d} \nn\\&&+ 160 
F_{bd}{}^{hij}{}_{,e} F_{cfhij,g} 
H_{a}{}^{ef,g} H^{abc,d}  - 160 
F_{bg}{}^{hij}{}_{,c} F_{dehij,f} 
H_{a}{}^{ef,g} H^{abc,d} \nn\\&&- 80 
F_{bc}{}^{hij}{}_{,e} F_{dfhij,g} 
H_{a}{}^{ef,g} H^{abc,d}  - 160 
F_{be}{}^{hij}{}_{,c} F_{dfhij,g} 
H_{a}{}^{ef,g} H^{abc,d} \nn\\&&- 160 
F_{be}{}^{hij}{}_{,c} F_{dghij,f} 
H_{a}{}^{ef,g} H^{abc,d}  - 240 
F_{bcd}{}^{hi,j} F_{efghj,i} 
H_{a}{}^{ef,g} H^{abc,d} \nn\\&&- 20 
F_{bc}{}^{hij}{}_{,g} F_{efhij,d} 
H_{a}{}^{ef,g} H^{abc,d}  - 120 
F_{bcd}{}^{hi,j} F_{efhij,g} 
H_{a}{}^{ef,g} H^{abc,d} \nn\\&& \left.+ 20 F_{bc}{}^{hij}{}_{,d} F_{efhij,g} 
H_{a}{}^{ef,g} H^{abc,d} + 480 F_{bcd}{}^{hi,j} F_{eghij,f} 
H_{a}{}^{ef,g} H^{abc,d}\nn\right).
\eea
Imposing the self-duality constraint also leads to a new coupling with structure $\e_{10}(\pa F^{(5)})^2 (\pa H)^2$ which has been neglected here.
 
\section{$\pa F^{(n)}\pa F^{(n-4)} (\pa H)^2$ couplings}

Since the minimum rank of RR field strength is $ 1 $, there is only one type of couplings in this section, \ie with $ n = 5 $. The effective action (\ref{IIB}) shows that the amplitude in the $s$ channel and contact terms vanishes. It also produces the following $ u $-channel amplitude
\bea 
A_u=\lp{\td V}_{F_1^{(n)}b_3C^{(n-3)}}\rp^{\mu_1\cdots \mu_{n-3}}\lp{\td G}_{C^{(n-3)}} \rp_{\mu_1\cdots \mu_{n-3}}{}^{\nu_1\cdots \nu_{n-3}}\lp{\td V}_{C^{(n-3)}F_2^{(n-4)}b_4}\rp_{\nu_1\cdots \nu_{n-3}}. \label{uFnFn-4HH}
\eea
where the vertices and propagator are given in \ref{FnbCn-3}, \ref{FnbCn+1} and \ref{RRpro}, respectively. The amplitude in the $t$ channel is the same as $A_u$ in which the particle labels of the external B-fields are interchanged which means that $A_t= A_u(3 \leftrightarrow 4)$. Gathering these two contributions together along with imposing the self-duality constraint, one finds the coupling
\bea
S_{\pa F^{(5)}\pa F^{(1)} (\pa H)^2} &=& -\frac{\gamma}{3^2.2^{5}\kappa^2} \int d^{10}x \ e^{2\phi_{0}} \sqrt{-G} \  (3 F_{cdefg,h} F^{a,b}
H_{a}{}^{cd,e} H_{b}{}^{fg,h} \nn\\&&+ 3 
 F{}{}_{cdfgh,b} F^{a,b}
H_{a}{}^{cd,e} H_{e}{}^{fg,h} -  
 F_{cdefg,h} F^{a,b}
H_{a}{}^{cd}{}_{,b} H^{efg,h})
\eea
plus a coupling of the form $ \e_{10}\pa F^{(5)}\pa F^{(1)} (\pa H)^2 $ which we drop it here. The above coupling is written in its minimal-term form. 
\section{$\pa F^{(n)}\pa F^{(n-2)} (\pa H) R$ couplings}

Here, we will calculate the couplings with structure $\pa F^{(n)}\pa F^{(n-2)} (\pa H) R$. There are three possibilities in this case, with $ n=3,4,5 $. The supergravity actions (\ref{IIA}) and (\ref{IIB}) dictate that the amplitudes in the $s$ channel and contact terms are given by
\bea 
A_s&=& 
\lp{\td V}_{F_1^{(n)}F_2^{(n-2)} b}\rp^{\mu \nu}\lp{\td G}_{b} \rp_{\mu \nu, \lambda\rho}\lp{\td V}_{b b_3 h_4}\rp^{\lambda\rho},\nonumber\\	A_c&=&{\td V}_{F_1^{(n)}F_2^{(n-2)} b_3 h_4}, \label{s-c-FnFn-2HR}
\eea
where the vertex operators and B-field propagator are given by: 
\bea
\lp{\td V}_{F_1^{(n)}F_2^{(n-2)} b}\rp^{\mu \nu}&=&- \frac{i\k}{(n-2)!}F_{1}^{\m\n}{}_{\nu_1\cdots\nu_{n-2}}F_2^{\nu_1\cdots\nu_{n-2}}, \label{FFb}
\eea 
\bea
\lp{\td G}_{b} \rp_{\mu \nu, \lambda\rho}&=&-\frac{i}{2 k^2}\left(\eta_{\mu\lambda}\eta_{\nu\rho}
-\eta_{\mu\rho}
\eta_{\nu\lambda}\right),\label{bpro}
\eea
\bea
\lp{\td V}_{b_1 h_2 b }\rp^{\m\n}&=& -2i \k \lp k_2.b_1.k_2 h_2^{\m \n}+2 k_2.b_1^{[\m} h_2^{\n]}.k_1+2k_2.b_1.h_2^{[\m}k_2^{\n]}\right. \nn\\&& \qquad\quad\left. +2k_1.k_2 h_2^{[\m}.b_1^{\n]}+k_2^{[\m} b_1^{\n]}.h_2.k_1+k_1.h_2.b_1^{[\m}k_2^{\n]}\rp,
\eea
\bea
{\td V}_{F_1^{(n)}F_2^{(n-2)} b_3 h_4}=\frac{i \k^2}{n!} \lp F_{1\mu_1 \cdots \mu_{n}}b_3^{[\m_1 \m_2} F_2^{\m_3 \cdots \m_{n}]}h_{4\l}{}^{\l}-2n F_{1\l\mu_2 \cdots \mu_{n}}b_3^{[\r \m_2} F_2^{\m_3 \cdots \m_{n}]}h_{4}{}^{\l}{}_{\r}\rp.
\eea
The amplitude  in the $u$ channel is the same as $A_s$ in which the particle labels of the RR $ (n-2) $-form field strength and B-field are interchanged, \ie $A_u= A_s(2 \leftrightarrow 3)$.
\bea 
A_u= 
\lp{\td V}_{F_1^{(n)}b_3 C^{(n-3)}}\rp^{\mu_1 \cdots \mu_{n-3}}\lp{\td G}_{C^{(n-3)}} \rp_{\mu_1 \cdots \mu_{n-3}}{}^{\nu_1 \cdots \nu_{n-3}}\lp{\td V}_{C^{(n-3)} F_2^{(n-2)} h_4}\rp_{\nu_1 \cdots \nu_{n-3}}. \label{uFnFn-2HR}
\eea

Similarly, the amplitude in the $t$ channel is the same as $A_u$ in which the particle labels of the external B-field and graviton are interchanged, \ie $A_t= A_u(3 \leftrightarrow 4)$.
\bea 
A_t= 
\lp{\td V}_{F_1^{(n)} h_4 C^{(n+1)}}\rp^{\mu_1 \cdots \mu_{n+1}}\lp{\td G}_{C^{(n+1)}} \rp_{\mu_1 \cdots \mu_{n+1}}{}^{\nu_1 \cdots \nu_{n+1}}\lp{\td V}_{C^{(n+1)} F_2^{(n-2)} b_3}\rp_{\nu_1 \cdots \nu_{n+1}}. \label{tFnFn-2HR}
\eea

Replacing the vertices and propagators in the above amplitudes and then summing them yields the following result for the couplings of one RR 3-form field strength, one RR 1-form field strength, one B-field strength and one Riemann tensor
\bea
S_{\pa F^{(3)}\pa F^{(1)} (\pa H) R} &=& -\frac{\gamma}{3.2^{5}\kappa^2} \int d^{10}x \ e^{2\phi_{0}} \sqrt{-G} \ \left(2 F^{cde,f} F^{a,b} H_{ac}{}^{g}{}_{,b} R_{dfeg} \right.\nn\\&&+ 4 F^{cde,f} F^{a,b} H_{ac}{}^{g}{}_{,f} R_{bdeg} - 10 F^{cde,f} F^{a,b} H_{af}{}^{g}{}_{,c} R_{bdeg}\nn\\&&- 2 F_{a}{}^{cd,e} F^{a,b} H_{bc}{}^{f,g} R_{dgef}+ 2 F_{a}{}^{cd,e} F^{a,b} H_{cd}{}^{f,g} R_{bgef} \nn\\&&- 2 F^{cde}{}_{,a} F^{a,b}
H_{cd}{}^{f,g} R_{bgef}- F_{a}{}^{cd}{}_{,b} F^{a,b}  
H_{c}{}^{ef,g} R_{dgef}\nn\\&&+ 4 F^{cde,f} F^{a,b} H_{cf}{}^{g}{}_{,a} R_{bdeg}+ 2 F^{cde,f} F^{a,b} H_{cf}{}^{g}{}_{,d} R_{aebg}\nn\\&& \left.+ F^{cde}{}_{,a} F^{a,b}
H_{c}{}^{fg}{}_{,b} R_{dfeg} \right)
\eea

The couplings of one RR 4-form field strength, one RR 2-form field strength, one B-field strength and one Riemann curvature can be written in the string frame as:
\bea
S_{\pa F^{(4)}\pa F^{(2)} (\pa H) R} &=&- \frac{\gamma}{3^2.2^{5}\kappa^2} \int d^{10}x \ e^{2\phi_{0}} \sqrt{-G} \ \left(12 F_{dfgh,e} F^{ab,c} H_{a}{}^{de,f} R_{bc}{}^{gh}\right. \nn\\&& - 6 F_{bdeh,g} F^{ab,c} H_{c}{}^{de,f} R_{a}{}^{g}{}_{f}{}^{h} - 12  F_{bdgh,e} F^{ab,c} H_{c}{}^{de,f} R_{a}{}^{g}{}_{f}{}^{h}\nn\\&&+ 6 F_{befh,g} F^{ab,c} H^{def}{}_{,c} R_{a}{}^{g}{}_{d}{}^{h} + 12 F_{begh,f} F^{ab,c} H^{def}{}_{,c} R_{a}{}^{g}{}_{d}{}^{h} \nn\\&&+ 6 F_{efgh,b} F^{ab,c} H^{def}{}_{,c} R_{a}{}^{g}{}_{d}{}^{h} - 12 F_{bcgh,f} F^{ab,c} H^{def,g} R_{a}{}^{h}{}_{de}\nn\\&&- 6 F_{befh,c} F^{ab,c} H^{def,g} R_{a}{}^{h}{}_{dg}- 12  F_{begh,f} F^{ab,c} H^{def,g} R_{a}{}^{h}{}_{cd}\nn\\&&- F_{cdef,h} F^{ab,c} H^{def,g} R_{abg}{}^{h}+ 3 F_{cdeh,f} F^{ab,c} H^{def,g} R_{abg}{}^{h}\nn\\&& - 6 F_{cefh,b} F^{ab,c} H^{def,g} R_{agd}{}^{h} + 12 F_{cefh,b} F^{ab,c} H^{def,g} R_{a}{}^{h}{}_{dg}\nn\\&& \left.- 2  F_{defh,b} F^{ab,c} H^{def,g} R_{a}{}^{h}{}_{cg}\right)
\eea

Finally, we find the couplings of one RR 5-form field strength, one RR 3-form field strength, one B-field strength and one Riemann curvature in the type IIB theory as:
\bea
S_{\pa F^{(5)}\pa F^{(3)} (\pa H) R} &=& -\frac{\gamma}{3^2.2^{5}\kappa^2} \int d^{10}x \ e^{2\phi_{0}} \sqrt{-G} \  \left(3 F_{bcefi,h} F^{abc,d} H_{d}{}^{ef,g} R_{a}{}^{h}{}_{g}{}^{i}\right.\nn\\&&+ 6 F_{bcehi,f} F^{abc,d} H_{d}{}^{ef,g} R_{a}{}^{h}{}_{g}{}^{i} - 12 F_{bcghi,f} F^{abc,d} H_{d}{}^{ef,g} R_{a}{}^{h}{}_{e}{}^{i}\nn\\&&- 12 F_{ceghi,f} F^{abc,d} H_{d}{}^{ef,g} R_{a}{}^{h}{}_{b}{}^{i} - 3 F_{bcfgi,h} F^{abc,d} H^{efg}{}_{,d} R_{a}{}^{h}{}_{e}{}^{i} \nn\\&&- 6 F_{bcfhi,g} F^{abc,d} H^{efg}{}_{,d} R_{a}{}^{h}{}_{e}{}^{i} - 6 F_{bfghi,c} F^{abc,d} H^{efg}{}_{,d} R_{a}{}^{h}{}_{e}{}^{i}\nn\\&&- 3 F_{bcfgi,d} F^{abc,d} H^{efg,h} R_{aeh}{}^{i}+ 3 F_{bcfgi,d} F^{abc,d} H^{efg,h} R_{ahe}{}^{i} \nn\\&&+ 6 F_{bcfhi,g} F^{abc,d} H^{efg,h} R_{a}{}^{i}{}_{de} + 12 F_{bdfgi,c} F^{abc,d} H^{efg,h} R_{aeh}{}^{i}\nn\\&&- 6 F_{bdfgi,c} F^{abc,d} H^{efg,h} R_{ahe}{}^{i} - 2   F_{befgi,c} F^{abc,d} H^{efg,h} R_{ahd}{}^{i}\nn\\&&\left.+ 12   F_{bfghi,c} F^{abc,d}  H^{efg,h} R_{a}{}^{i}{}_{de} + 4 F_{defgi,c} F^{abc,d} H^{efg,h} R_{ahb}{}^{i} \right)
\eea

Similar to the previous cases, imposing the self-duality constraint in the above action leads to a new coupling with structure $\e_{10} \pa F^{(5)}\pa F^{(3)} (\pa H) R$. We also find out that there is another coupling with structure $\e_{10}\pa F^{(n)}\pa F^{(n-2)} (\pa H) R$ which have been not considered here.

\section{$\pa F^{(n)}\pa F^{(n-2)} \pa \ph \pa H$ couplings}

The last coupling that we aim to obtain is $\pa F^{(n)}\pa F^{(n-2)} \pa \ph \pa H$. There are three types of couplings in this case and those are with $ n=3,4,5 $. When $ n=3,4 $, from the supergravity actions (\ref{IIA}) and (\ref{IIB}), one can observe that the field-theory amplitudes in the $s$ channel and contact terms are given by
\bea 
A_s&=& 
\lp{\td V}_{F_1^{(n)}F_2^{(n-2)} b}\rp^{\mu \nu}\lp{\td G}_{b} \rp_{\mu \nu, \lambda\rho}\lp{\td V}_{b \phi_3 b_4}\rp^{\lambda\rho},\nonumber\\	A_c&=&{\td V}_{F_1^{(n)}F_2^{(n-2)} \phi_3 b_4}, \label{s-c-FnFn-2DH}
\eea
respectively. Here, the vertices that have not been previously introduced are given by: 
\bea
\lp{\td V}_{\phi_1 b_2 b}\rp^{\m\n}&=&-\sqrt{2}i\k\lp 2 k_1.b_2^{[\n} k_2^{\m]}-k_1.k_2 b_2^{\m \n}\rp,
\eea
\bea
{\td V}_{F_1^{(n)}F_2^{(n-2)} \phi_3 b_4}&=&\frac{\sqrt{2}i \k^2(5-n)}{n!}F_{1\mu_1 \cdots \mu_{n}}b_4^{[\m_1 \m_2} F_2^{\m_3 \cdots \m_{n}]}.
\eea
The $u$-channel amplitude $A_u$ can now be obtained from the $A_s$ by permuting the particles lines $ 2 $ and $ 3 $ Similarly, the $t$-channel amplitude $A_t$ is obtained by permuting the external lines $ 3 $ and $ 4 $
\bea 
A_u= 
\lp{\td V}_{F_1^{(n)}\phi_3 C^{(n-1)}}\rp^{\mu_1 \cdots \mu_{n-1}}\lp{\td G}_{C^{(n-1)}} \rp_{\mu_1 \cdots \mu_{n-1}}{}^{\nu_1 \cdots \nu_{n-1}}\lp{\td V}_{C^{(n-1)} F_2^{(n-2)} b_4}\rp_{\nu_1 \cdots \nu_{n-1}}. \label{uFnFn-2DH}
\eea

Similarly, the $t$-channel amplitude $A_t$ is obtained by permuting the external lines $ 3 $ and $ 4 $
\bea 
A_t= 
\lp{\td V}_{F_1^{(n)} b_4 C^{(n-3)}}\rp^{\mu_1 \cdots \mu_{n-3}}\lp{\td G}_{C^{(n-3)}} \rp_{\mu_1 \cdots \mu_{n-3}}{}^{\nu_1 \cdots \nu_{n-3}}\lp{\td V}_{C^{(n-3)} F_2^{(n-2)} \phi_3}\rp_{\nu_1 \cdots \nu_{n-3}}. \label{tFnFn-2DH}
\eea

The sum of the pole diagrams, after a simple calculation, leads to the following couplings for one RR 3-form field strength, one RR 1-form field strength, one dilaton and one B-field strength in the type IIB supergravity
\bea
S_{\pa F^{(3)}\pa F^{(1)} \pa \ph \pa H} &=& \frac{\gamma}{3.2^{\frac{9}{2}}\kappa^2}  \int d^{10}x \ e^{2\phi_{0}} \sqrt{-G} \left(  F^{cde,f} F^{a,b} \phi_{a,b} H_{cdf,e}  \right. \nn\\&&  - 3
F^{def}{}_{,b} F^{a,b} \phi_{a}{}^{,c} H_{cde,f}-F_{c}{}^{de,f} F^{a,b} \phi_{a}{}^{,c} H_{def,b} \nn\\&& \left.+ 2 F_{a}{}^{ef}{}_{,b} F^{a,b} 
\phi^{c,d} H_{cef,d} 
 - 2 F_{a}{}^{ef}{}_{,c} F^{a,b} \phi^{c,d} H_{def,b} 
\right).\label{F3F1DH}
\eea

The amplitude of one RR 4-form field strength, one RR 2-form field strength, one dilaton and one B-field strength produces the couplings
\bea
S_{\pa F^{(4)}\pa F^{(2)} \pa \ph \pa H} &=&  \frac{\gamma}{3^2.2^{\frac{13}{2}}\kappa^2}  \int d^{10}x \ e^{2\phi_{0}} \sqrt{-G} \  \left(12  F_{cefg,d} F^{ab,c}  
\phi_{a}{}^{,d} H_{b}{}^{ef,g} \right.\nn\\&&+ 12 
F_{defg,c} F^{ab,c} \phi_{a}{}^{,d} H_{b}{}^{ef,g} - 12 
F_{bdef,g} F^{ab,c} \phi_{a}{}^{,d} H_{c}{}^{ef,g} 
\nn\\&&+ 12 F_{bdeg,f} F^{ab,c} \phi_{a}{}^{,d} H_{c}{}^{ef,g} 
- 2 F_{aefg,b} F^{ab,c} \phi_{c}{}^{,d} H^{efg}{}_{,d} 
 \nn\\&&+ 12 
F_{bdfg,e} F^{ab,c} \phi^{d,e} H_{c}{}^{fg}{}_{,a} 
 - 9 F_{abfg,e} F^{ab,c} \phi^{d,e} H_{d}{}^{fg}{}_{,c} 
 \nn\\&& \left.+ 24
F_{aefg,b} F^{ab,c} \phi^{d,e} H_{d}{}^{fg}{}_{,c} 
+ 6 F_{acfg,b} F^{ab,c} \phi^{d,e} H_{d}{}^{fg}{}_{,e} 
\right)\label{F4F2DH}
\eea
in the string frame. In particular, when $ n=5 $, as can be seen from type IIB supergravity action (\ref{IIB}), the $ u $-channel amplitude as well as contact terms vanishes. On the other words, the $ s $ and $ t $ channels are the only two diagrams that contribute to the process. Putting these two contributions together, one can obtain the action which reproduce the corresponding amplitude as
\bea
S_{\pa F^{(5)}\pa F^{(3)} \pa \ph \pa H} &=& \frac{\gamma}{3^2.2^{\frac{11}{2}}\kappa^2}  \int d^{10}x \ e^{2\phi_{0}} \sqrt{-G} \  \left(2  F_{cdegh,b} F^{cde,f} \phi^{a,b} H_{af}{}^{g,h} \right. \nn\\&&+ F_{cdegh,b} F^{cde,f}  \phi^{a,b} H_{f}{}^{gh}{}_{,a}  - 3 F_{adegh,b} F^{cde,f} \phi^{a,b} H_{f}{}^{gh}{}_{,c} \nn\\&& \left.+ 4 F_{bcfgh,d} F_{a}{}^{cd,e} \phi^{a,b} H^{fgh}{}_{,e} + F_{cdfgh,b} F_{a}{}^{cd,e} \phi^{a,b} H^{fgh}{}_{,e} \right).\label{F5F3DH}
\eea

The number of terms in (\ref{F3F1DH}) and (\ref{F5F3DH}) has been reduced as much as possible. Here also imposing the self-duality condition yields a coupling with structure $\e_{10} \pa F^{(5)}\pa F^{(3)} \pa \ph \pa H $ which is not considered here because the number of indices is too large to find all possible contractions of this structure.

We have compared the results of this paper with the corresponding couplings which have been obtained from string amplitude calculations in RNS \cite{Bakhtiarizadeh:2013zia} and pure-spinor formalisms \cite{Policastro:2006vt} as well as T-duality approach \cite{Garousi:2013nfw,Bakhtiarizadeh:2013zia}, and find an exact agreement when we write both couplings in terms of independent variables.  

It is interesting to find the remaining couplings by improving the computer program. One can also apply the S and T-duality approaches to get these couplings.   

\section*{Acknowledgement}\addcontentsline{toc}{section}{Acknowledgement}

This work has been financially supported by the research deputy of Sirjan University of Technology.


\providecommand{\href}[2]{#2}\begingroup\raggedright
\endgroup
\end{document}